# Piezoelectric Domains in the AlGaN Hexagonal Microrods: Effect of Crystal Orientations

A. K. Sivadasan,[1,a)] G. Mangamma,[2,a)] Santanu Bera,[3] M. Kamruddin,[2] and Sandip Dhara[1,a)]

[1] Nanomaterials and Sensor Section, Surface and Nanoscience Division, Indira Gandhi Centre for Atomic Research, Kalpakkam-603102

[2] Nanomaterials Characterization Section, Surface and Nanoscience Division, Indira Gandhi Centre for Atomic Research, Kalpakkam-603102

[3] Water and Steam Chemistry Laboratory, BARC Facilities, Kalpakkam-603102, India

[a)] Authors to whom correspondence should be addressed. Electronic addresses: sivankondazhy@gmail.com, gm@igcar.gov.in and dhara@igcar.gov.in

*Abstract*

Presently, the piezoelectric materials are finding tremendous applications in the micro-mechanical actuators, sensors and self-powered devices. In this context, the studies pertaining to piezoelectric properties of materials in the different size ranges are very important for the scientific community. The III-nitrides are exceptionally important, not only for optoelectronic but also for their piezoelectric applications. In the present study, we synthesized AlGaN via self catalytic vapor-solid mechanism by atmospheric pressure chemical vapor deposition technique on AlN base layer over intrinsic Si(100) substrate. The growth process is substantiated using X-ray diffraction and X-ray photoelectron spectroscopy. The Raman and photoluminescence study reveal the formation of AlGaN microrods in the wurtzite phase and ensures the high optical quality of the crystalline material. The single crystalline, direct wide band gap and hexagonally shaped AlGaN microrods are studied for understanding the behavior of the crystallites under the application of constant external electric field using the piezoresponse force microscopy. The present study is mainly focused on understanding the behavior of induced polarization for the determination of piezoelectric coefficient of AlGaN microrod along the *c*-axis and imaging of piezoelectric domains in the sample originating because of the angular inclination of AlGaN microrods with respect to its AlN base layers.

## I. INTRODUCTION

Group III-nitrides and their ternary and quaternary alloys with wide-direct band gap have immense attention in scientific research for their different applications in the semiconducting devices including white-blue diodes and lasers.[1] Engineering the band gap of AlN, GaN and InN finds different applications in the semiconducting industry for their various optoelectronic properties such as photovoltaic, hydrogen storage, light and field emissions.[2-6] Because of the capability to create two-dimensional electron gas at the hetero-junctions, the III-nitrides are used for developing high electron mobility transistors along with the hetero-junction field effect transistors and bipolar transistors.[7] III-nitride based photonic device is also used as ultraviolet (UV)-blue light emissions.[2,8] Among all the III-nitrides, AlGaN is very important because of its applications in high power and high mobility electronic devices.[9,10] The major advantages of the AlGaN alloys are its compatibility of tuning the direct band gap in between the energy ranges from 3.47 to 6.2 eV. Thus, AlGaN ternary alloys with different composition plays an important role in UV and deep-UV optoelectronic device applications.[2,11]

In addition to their utility as direct band gap material, the III-nitrides exhibit piezoelectricity which establishes their importance, particularly, in different micro-mechanical devices such as micro-actuators, micro-sensors, ultrasonic motors, micro-pumps, self-powered devices, electro-mechanical sensing, mechanical energy harvesting and other various applications as MEMS based device.[12-16] Piezoelectricity is a phenomenon in which an electric field is generated inside a material subjected to a mechanical strain or vice versa.[17,18] It implies that the piezoelectric effect can be employed to convert mechanical energy into electrical energy also. Electromechanical coupling characteristics of various ferroelectric and piezoelectric materials can be studied using piezo-response force microscopy (PFM).[19,20] The non-centrosymmetric wurtzite III-nitrides are known to posses high piezoelectric co-efficient which leads to the polarization of ions in a crystal.[21] The space group of wurtzite crystal structure is *P6₃mc*, with polar in nature and it shows spontaneous polarization.[22] Therefore, a piezoelectric potential may be generated across the non-centrosymmetric crystal by applying an external stress or vice versa.[23] Consequently, this strained material behaves like a charged capacitor with an electrostatic potential across it, which can be utilized for the above mentioned applications.[17,20]

In this report, we present the PFM imaging of single crystal AlGaN hexagonal microrods on AlN base layer to investigate the piezoelectric domains as well as piezoelectric coefficient along with the detailed analysis for the growth process using X-ray diffraction (XRD) and X-ray photoelectron spectroscopy (XPS). Raman and photoluminescence (PL) spectroscopic studies are used to confirm the formation of wurtzite phase and optical quality of the sample, respectively. The PFM imaging of crystals is used to examine the piezoelectrically induced displacement of the surface layers of the materials which are actuated by an external AC modulation voltage.

## II. EXPERIMENTAL

### A. Synthesis and characterizations

AlGaN microrods were synthesized using a atmospheric pressure chemical vapour deposition (APCVD) technique with high pure Ga metal drop (99.999%, Alfa Aesar) and Al coated (75 nm) intrinsic Si(100) substrate as precursors for Ga and Al vapor, respectively. Al ingot (99.999%, Alfa Aesar) was used for coating Al on intrinsic Si(100) substrate in the thermal evaporation technique with a base pressure of $1\times10^{-6}$ mbar. Ultra high pure ammonia ($NH_3$, 99.999%) was used as reactive gas with a continuous and constant flow rate of 50 sccm and the growth was carried out at 1100 $^o$C for 2h.

The vibrational characterization was performed using Raman spectroscopy (inVia, Reinshaw) with 514.5 nm excitation of an $Ar^+$ laser. A grating with 1800 gr.mm$^{-1}$ was used for monochromatization of the scattered waves. A thermoelectrically cooled CCD detector was used for recording the spectra in the backscattering geometry. The spectra were collected with the help of 100X objective with numerical aperture (N.A.) value of 0.85. To investigate the band gap features and the optical quality of the samples, the AlGaN microrods were excited with a UV laser of wavelength 325 nm (He-Cd laser) and the photoluminescence (PL) spectra were recorded at room temperature with 2400 gr.mm$^{-1}$ grating as the monochromatizer using the same detector as used for the Raman characterization. The PL spectra were collected with the help of 40X micro-spot NUV objective with an N.A.value of 0.50. The structural confirmation for AlN base layer was performed using X-ray diffractometer (Brucker, D8 Discover). The X-ray photoelectron spectroscopic (XPS; VG ESCALAB MK200X) analysis were for both the AlGaN microrods as well as AlN base layers using an X-ray source of Al-K$\alpha$ (1486.6 eV), and the binding energy values were measured with respect to the C $1s$ reference peak. The X-ray beam diameter used for the XPS measurements is around 3 mm and the collection area (with the largest slit) for the analysis is approximately $2 \times 3$ mm$^2$. The spectra were processed by applying Shirley type background and curve fitted with mixture of Gaussian−Lorentzian line shapes.

## B. Piezoresponse force microscopic (PFM) set up

For measuring the piezoelectric properties of the ferroelectric or piezoelectric materials using contact mode PFM, a conductive atomic force microscopic (AFM) tip was employed. The schematic representation of the PFM measurement set up is shown in the figure 1, and its working procedure includes two main steps. The first step was the polarization in a predefined area of the sample by using a DC bias voltage, and the second step was detecting the magnitude of the piezoelectric response by means of AC electric fields. In the second step, to detect the magnitude of the induced piezoelectric response, the AFM tip was connected to a lock-in amplifier with PFM configuration. An external DC ($V_{dc}$) bias and an AC voltage [$V_{ac}cos(\omega t)$] was then applied between the cantilever and the grounded substrate for activating the piezoelectric domains and probing the feedback, respectively.

$$V = V_{dc} + V_{ac}\ cos(\omega t) \quad\text{----------------------------------------------------- (1)}$$

As a result, cantilever displacement occurs due to piezoelectric effect.

$$\Delta Z = \Delta Z_{dc} + \Delta Z_{ac}\ cos(\omega t + \varphi) \quad\text{------------------------------------------- (2)}$$

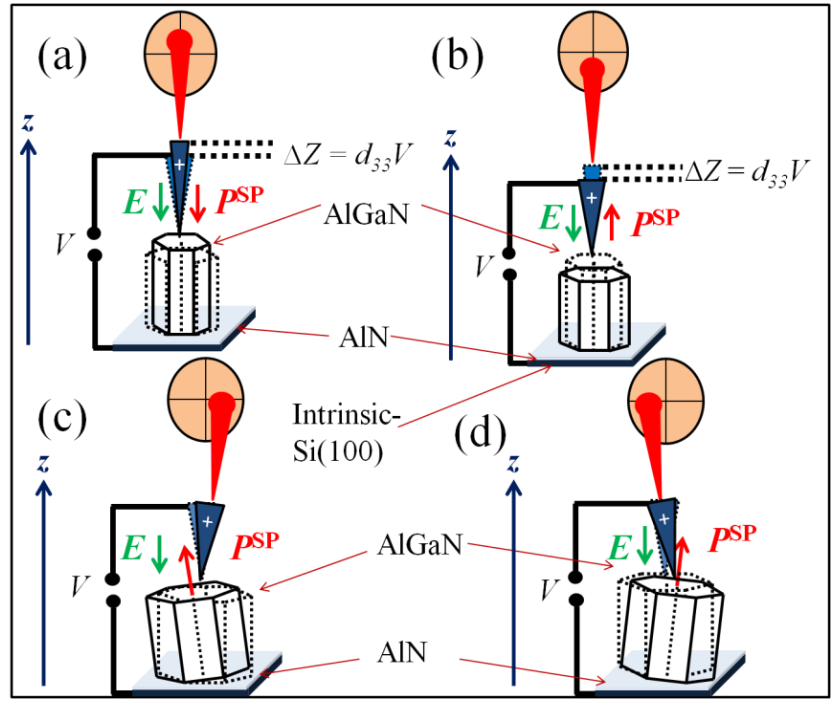

FIG. 1. The schematic representation of PFM imaging set up for different polarization conditions and crystalline orientations. (a) $E$ is parallel to $P^{SP}$ (b) $E$ is anti-parallel to $P^{SP}$ (c) and (d) $E$ is making an angle with $P^{SP}$.

By applying the voltage between the sample surface and the AFM tip, an external electric field was setup across the sample. The sample would locally enlarge or contract according to the electric field due to the electrostriction, as an "inverse piezoelectric" effect. When the initial spontaneous polarization ($P^{SP}$) of the electrical domain of the measured sample is perpendicular to the sample surface, and parallel to the applied electric field ($E$); the domains experience a vertical expansion. Since the AFM tip is in contact with the sample surface, the increase in domain size bends the AFM cantilever and pushes it up. It results in the increase of tip deflection compared to the position before applying the electric field [Fig 1(a)]. On the contrary, the domain contracts and in turn results in a decreased cantilever deflection if the initial domain polarization is anti-parallel to the applied electric field [Fig 1(b)]. The magnitude of change in cantilever deflection, in such situation, is directly related to the amount of expansion or contraction of the sample electric domains, and hence it is proportional to the applied electric field.[19,20] Thus, the amplitude of the PFM response is proportional to the strength of the piezo-response and the phase signal is the representative of the polarization direction of the samples below the probe tip [Figs. 1(b) and 1(c)]. In the present study, a predefined 10 × 10 µm² area was selected for PFM imaging and an external bias voltage of +5 V and -5V was applied between the tip and the bottom substrate. A diamond like carbon (DLC) coated stiff cantilever with a resonant frequency of 256 KHz and a stiffness constant of 11.5 N/m used as an AFM tip (with a cantilever dimension of 100 × 35 × 2 µm) in the contact mode for inducing the polarization in AlGaN microrods. The deformations in the crystal surfaces are detected by the AFM tip through the optical laser beam path and with the help of a four-quadrant photodetector. The lock-in amplifier de-convolutes the induced signal on the cantilever to measure the amplitude and phase difference ($\varphi$) with respect to the input AC voltage.

## III. RESULTS AND DISCUSSIONS

### A. The morphological analysis

The AFM images of the samples show (Fig. 2) hexagonal shaped microrods with an average diameter of 2-5 µm and an average length of 5-10 µm. The AlGaN single crystal hexagonal microrods are showing perfect planar and morphologically smooth facets. For understanding the statistical distribution of microrods over the substrate, we recorded another topographic image from different portion of the sample used for the PFM study (supplementary Fig. S1).[24] The corresponding line profiles of AlGaN mocrorods are also shown in the supplementary figure S2.[24] The root mean square (RMS) and average size distribution of AlGaN microrods over the sample are found to be 0.634 and 0.440 µm, respectively.

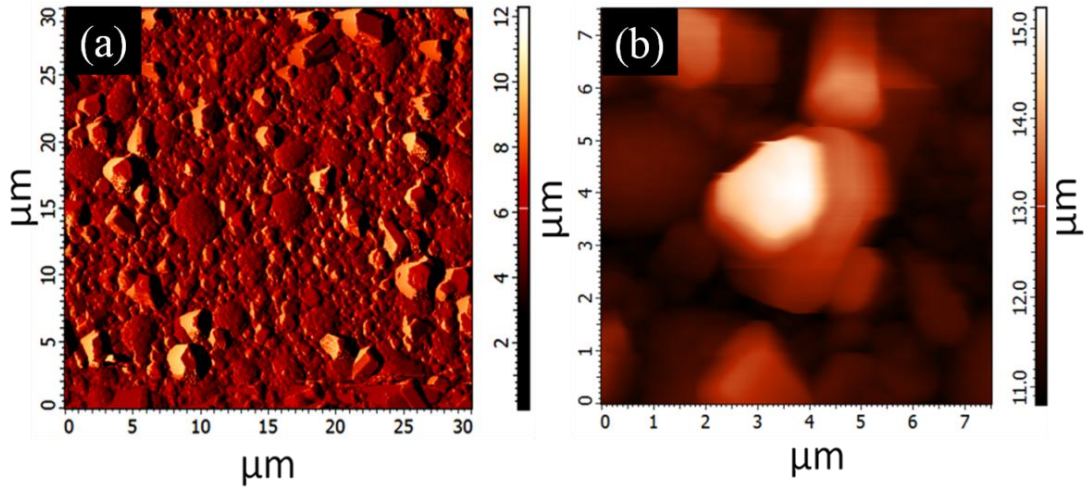

FIG. 2. (a) Low and (b) high magnification images of the AFM topographic features of AlGaN hexagonal microrods.

As the AFM tip does not allow the measurement of abrupt variation of height correctly, the AFM images in figure 2 are not always the most suitable technique to prove the shape and size of the microrods accurately. Therefore, we have collected FESEM images in order to show the shape of microrods directly and to measure their actual size and height. The FESEM image of as-prepared sample shows the perfect hexagonal shaped microrods with an average diameter of 2-5 µm and an average length of 5-10 µm which are similar to that measured from AFM images (Fig. 3). Inset of the figure 3 shows the magnified image of a single microrod with perfect planar and morphologically smooth facets. The average number of crystallites is found out to be $25 \times 10^3$ per mm$^2$, as analysed from FESEM image.

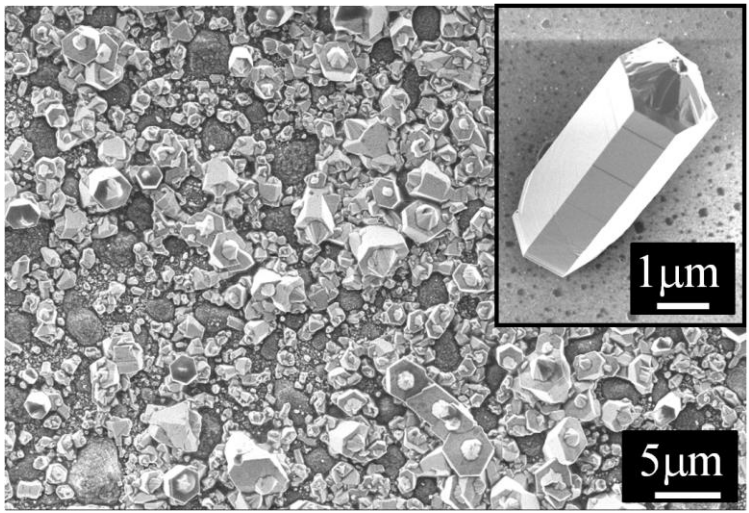

FIG. 3. The FESEM images of AlGaN hexagonal microrods. Inset showing the magnified image of a single microrod.

**B. The vapor-solid (VS) growth mechanism**

The growth process of AlGaN microrods involves a self catalytic vapor-solid (VS) mechanism including two concomitant steps, as follows. In the first step, the Al thin film, coated on the intrinsic Si(100) substrate, may react with atomic N at 1100 $^oC$ to form very thin AlN islands (base layer) which may act as reactive seed for promoting the growth process. In the second step, the un-reacted portions in the Al thin film supply the atomic Al for the further growth process.

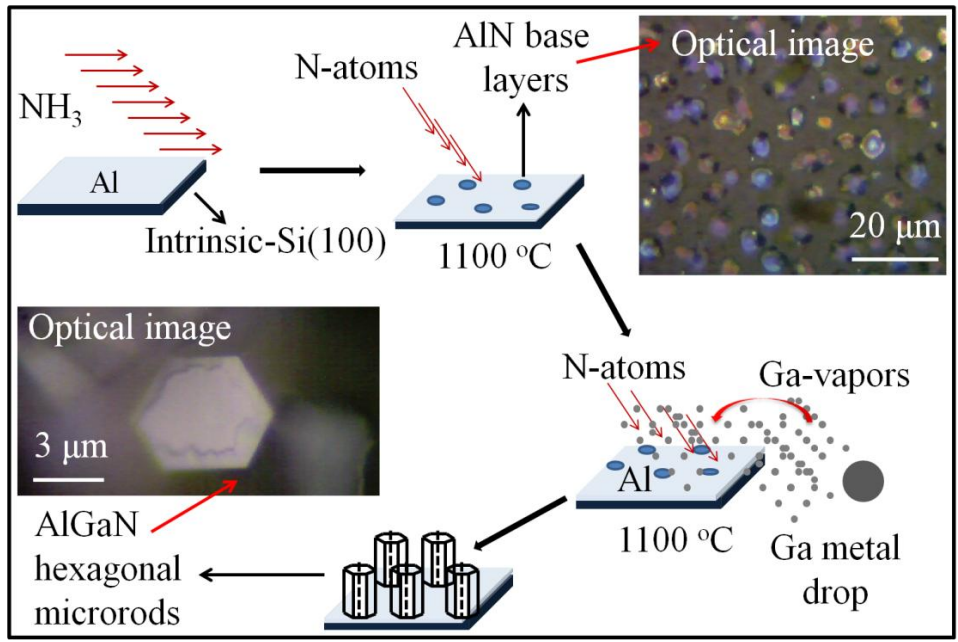

FIG. 4. The schematic growth process of AlGaN microrods in self catalytic vapor-solid mechanism.

The proposed growth mechanism for AlGaN microrod via vapor-solid (VS) growth process is depicted schematically in the figure 4. The Ga metal precursor as well as Al thin film was kept in the single zone furnace at an optimized growth temperature of 1100 °C and the ultra high pure $NH_3$ with a constant flow rate of 50 sccm was allowed to flow in to the reaction chamber. At this temperature, the high pure $NH_3$ was decomposed to form reactive atomic N and H (vapor phase). Consequently, the atomic N reacted to some portions on the surface layers of Al thin film and forms few islands of AlN base layers over the intrinsic Si(100) substrate. The optical image of such AlN base layer is shown in the right top inset of figure 4. Continued supply of the metallic Ga vapors and the Al evaporated from the un-reacted portions of the thin film lead to the further growth process of AlGaN microrods (solid phase) without the aid of any metallic catalytic particles. The optical image of the as grown AlGaN microrod is also shown in the left bottom inset of figure 4. Therefore, the AlGaN microrods grown on the substrate via VS growth mechanism appeared to be well separated and which may possess AlN islands as a base layer.[25-30] Typical morphological, vibrational and optical studies on different samples are included in the supplementary figures S3, S4 and S5, respectively.[24]

## C.  The structural and vibrational analysis

Crystallographic structural study of the fully grown substrate with microrods show [Fig. 5(a)] (*hkl*) planes of (100), (002), (101), (102), (110), (103), (200), (112), (201), (004) and (202) at 2θ values of 32.41, 34.53, 36.83, 48.10, 57.77, 63.44, 67.75, 69.10, 70.50, 72.94 and 78.45 degree, respectively, which corresponding to the GaN (JCPDS # 00-050-0792). The absence of any peak shift due the presence of Al may be due to the low Al content in sample. Polycrystalline nature in the XRD pattern indicates the presence of all possible orientation in ensemble of these microrods. The presence of AlN base layers over the intrinsic Si(100) substrate is confirmed by the XRD analysis [Fig. 5(b)] of annealed AlN thin film at 1100°C for 1 hr in the atomic N atmosphere. This additional experiment was performed as it was difficult for identifying AlN base layer in the fully grown substrate with the large amount of microrods. The crystallographic structural study shows that (*hkl*) planes of (200) and (220) at 2θ values of 45.76 and 67.08 degree, respectively, which correspond to the cubic phase of AlN (JCPDS # 00-046-1200).

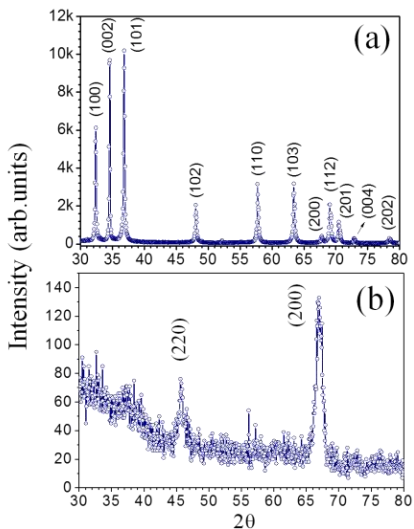

FIG. 5. The XRD pattern of (a) microrod in the wurtzite GaN phase and (b) cubic AlN phase of base layers on intrinsic Si(100) substrate.

In order to confirm the phase of the as-prepared hexagonally shaped micro-crystals, we carried out the vibrational studies using Raman spectroscopy (Fig. 6) by focusing the laser beam with a spot size of 1 μm, very precisely and exactly on a single and isolated microrod (size ~3 μm). Group theoretically, the allowed Raman modes for hexagonal wurtzite structures such as InN, GaN and AlN at the $\Gamma$ point are $\Gamma_{acoustic}+ \Gamma_{optical} = (A_1+E_1) + (A_1+2B_1+E_1+2E_2)$. Among the total eight allowed Raman modes, one set of $A_1$ and $E_1$ modes are acoustic and remaining all other six modes are originated due to the optical phonons. Another set of $A_1$ and $E_1$ modes in the optical phonon group are both Raman and infrared (IR) active. The $E_2$ modes are also optical phonons but they are only Raman active, and the $B_1$ modes are silent. Because of the polar nature of the unit cell, $A_1$ and $E_1$ modes can again split into longitudinal optical (LO) and transverse optical (TO) modes. Therefore, there are six optical modes of $A_1(TO)$, $A_1(LO)$, $E_1(TO)$, $E_1(LO)$, $E_2^L$ and $E_2^H$, which can be active in the first-order Raman scattering of wurtzite structured polar materials.[31-33]

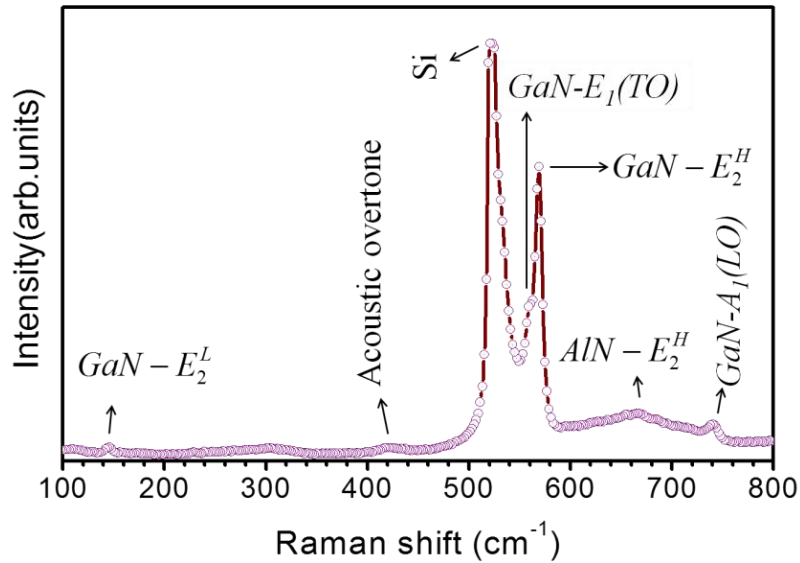

FIG. 6. Typical Raman spectra of an AlGaN single crystal hexagonal microrod.

The spectral lines centered at 146, 425, 559, 569 and 741 cm$^{-1}$ (Fig. 6) are assigned to allowed symmetric Raman modes of $E_2^L$, Acoustic overtone, $E_1(TO)$, $E_2^H$ and $A_1(LO)$ modes of wurtzite GaN phase, respectively.[32] The presence of an extra peak centered at 667 cm$^{-1}$ is assigned as AlN-$E_2^H$ mode. It is reported that, the Raman $A_1(TO)$, $E_1(TO)$, $E_2^L$ and $E_2^H$ phonon modes obeys two-mode behavior of random alloy, meanwhile the $E_1(LO)$ and $A_1(LO)$ modes follows the one-mode phonon behavior.[31,33] The observation of GaN- $E_2^H$ mode along with the AlN- $E_2^H$ in a single Raman spectrum recorded from a single hexagonal micro-crystal indicates the two-mode behavior of the phonons in the random alloy formation of the AlGaN phase.[31,32] Moreover, we observed a significant blue shift in $A_1(LO)$ mode of the AlGaN crystal (741 cm$^{-1}$) as compared to that for the pure GaN (~730 cm$^{-1}$). The percentage of Al in the AlGaN random alloy is estimated to be 3 at % using the formula $A_1(LO) = 734 + 153x - b_{A1(LO)}\ x(1-x)$ with the bowing parameter of $b_{A1(LO)} = -75$ cm$^{-1}$.[31] In order to understand the compositional homogeneity and crystalline nature of the pristine AlGaN microrod, we carried out a Raman imaging of the integrated intensity distribution of $E_2^H$ mode along one isolated crystal (supplementary Fig. S6).[24] The Raman imaging results supports the high quality crystalline and compositionally homogenous nature of the microrod.

### D. The elemental analysis: X-ray photoelectron spectroscopy

We carried out the X-ray photoelectron spectroscopy (XPS) for the bare AlN thin film as well as AlGaN microrods grown on AlN base layer, for further elemental analysis of the respective samples. The Shirley type back ground corrected XPS spectra for different elements and their characterisitic electronic transitions for each sample is shown in the figure 7.

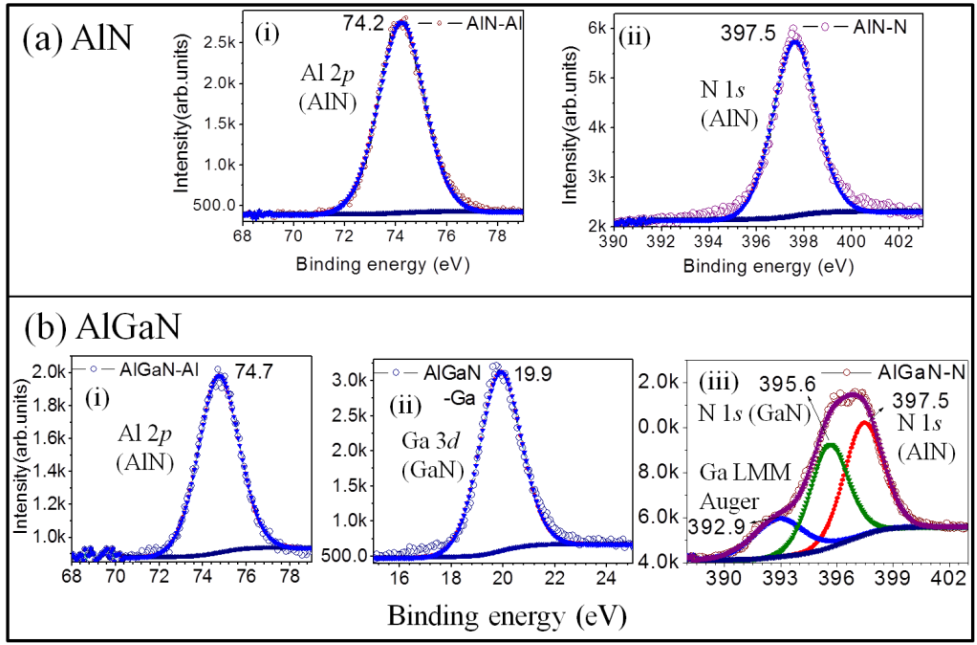

FIG. 7. Typical XPS spectra for different elements and its characteristic electronic transitions of (a) AlN thin film and (b) AlGaN hexagonal microrods.

In the case of AlN thin film [Fig. 7(a)], the binding energy for Al 2p (AlN) level is observed at 74.2 eV [Fig. 7a (i)] and N 1s (AlN) core level is identified at 397.5 eV [Fig. 7a (ii)]. Similarly, in the case of AlGaN microrods grown on the AlN base layer [Fig. 7(b)], the Al 2p level transition is observed at 74.7 eV [Fig. 7b (i)]. In this sample, we can expect that the XPS signal may show the cumulative effect of AlN base layer as well as from the AlGaN microrods. However, the presence of an additional peak centered at 19.9 eV corresponds to Ga 3d (GaN) level transitions [Fig. 7b (ii)] in the AlGaN sample compared to the AlN thin film, further confirms the incorporation of Ga along with Al in the sample. The peak positions of both Al 2p and Ga 3d are well matching with the reported literature values for AlGaN samples.[35] N 1s spectrum of AlGaN sample is very broad along with the presence of Ga Auger peak. N 1s peak is deconvoluted into two distinct peaks [Fig. 7b (iii)]; the deconvoluted peak observed at 397.3 eV is due to the contribution from the N 1s (AlN) level transitions and the other deconvoluted peak centered at 395.6 eV appear due to the N 1s from GaN or AlGaN level transitions. The peak at around 392.9 eV in N 1s spectrum is due the Ga LMM Auger transition.[34-37] Therefore, the XPS study provides a supportive and substantiating evidence for the incorporation of Al in the proposed VS growth mechanism. The mean free path of the photoelectrons Al 2p, N 1s and Ga 3d is approximately 2 nm and the corresponding depth information is around 6 nm. However, the XPS spectra were recorded from the whole area of the sample, so the spectral intensity was influenced, not only from the microrods but also from the AlN base layer. Therefore we cannot estimate the Al content of single AlGaN microrod from the above spectra.

**E. Confirmation of Al in the as prepared AlGaN microrods: Photoluminescence spectroscopy**

Room temperature photoluminescence (PL) spectrum (Fig. 8) of a single AlGaN hexagonal microrod shows strong emission peaks centered at 3.52 and 3.32 eV.

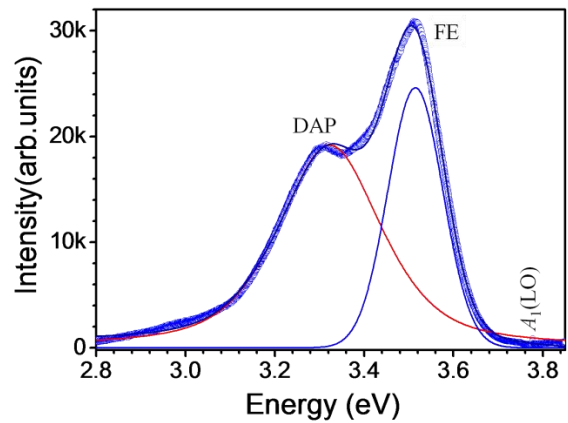

FIG. 8. Typical photoluminescence spectra of an AlGaN single crystal hexagonal microrod.

The emission peak centered at 3.52 eV is due to the recombination of free exciton (FE) from the conduction band minimum to the valence band maximum. The luminescence peak observed at 3.32 eV is originated because of the recombination of the neutral donor-acceptor pair (DAP; $D^0A^0$). This luminescence peak observed due to a transition from a shallow donor state of nitrogen vacancy ($V_N$) to a deep acceptor state of Ga vacancy ($V_{Ga}$).[32] The significant blue shift in the band edge emission compared to pure GaN (~ 3.47 eV) is also a strong indication for the presence of Al in as-prepared sample. From the PL shift of AlGaN microrods compared to that of pure GaN, the approximate percentage of Al in the as grown sample is found to be ~ 2 at% using the formulation of band bowing with respect to the observed and stoichiometric AlN ($E_g$ ~ 6.2 eV) and GaN ($E_g$ ~3.47 eV) phases. So the value of the Al content, as derived from both Raman and PL spectroscopic techniques, matches fairly with each other. A very high PL intensity obtained in these single crystals even for low excitation laser power ensures the high optical quality of the crystals.

**F. Piezoresponse force microscopic (PFM) imaging of AlGaN microrods**

*1. The domain switching in the piezoelectric zones of AlGaN microrod*

Piezoelectricity linearly relates an external electrical field to mechanical strain of the material, leading to a direct piezoelectric effect ($P_i = d_{ijk} e_{jk}$) and a converse piezoelectric effect ($\varepsilon_{ij} = d_{ijk} E_k$), where $P$ is the polarization vector, $e_{jk}$ is the stress tensor, $\varepsilon_{ij}$ is the strain tensor, $E$ is the electric field vector, and $d_{ijk}$ is the piezoelectric tensor with a rank three.[20,38] For the AlGaN/AlN system, the total polarization $P$ arises particularly from two sources, the first one from the spontaneous polarization (SP) originated due to the nature of the crystal and the polarity of the interfaces between the hetero-junctions, and the other one is piezoelectric (PE) polarization due to the lattice strain. The spontaneous electric polarization ($P^{SP}$) and piezoelectric polarizations ($P^{PE}$) generate a built-in electrostatic field all over the sample surface.[16] Eventually, both of the above discussed polarizations and its discontinuity at the inter-junctions lead to a net charge density ($\sigma$).[7] This net charge density and immobile piezoelectric polarization charges may play a major key role in the interaction of the wurtzite-structured AlGaN with the external electric field as well as the mechanical strain.[16] For AlGaN micro-crystals, the spontaneous polarization is given by,

$$\vec{P}^{SP} = P^{SP}.\hat{z} \quad\quad\quad (3)$$

Where $\hat{z}$ is a unit vector in the [0001] direction.[7] For any wurtzite structured materials such as InN, AlN, GaN, and its alloys, *e.g.* AlGaN, the induced piezoelectric polarization field can be written as follows,[7]

$$P^{PE} = (d_{51}\varepsilon_{31})\hat{x} + (d_{51}\varepsilon_{23})\hat{y} + (d_{31}\varepsilon_{11} + d_{31}\varepsilon_{22} + d_{33}\varepsilon_{33})\hat{z} \quad \text{------------------------ (4)}$$

Taking z along the c-axis, $\varepsilon_{11}$, $\varepsilon_{22}$ and $\varepsilon_{33}$ are the strains along the *x*, *y* and *z*-axis, respectively, and $\varepsilon_{23}$ and $\varepsilon_{31}$ are shear strains. Here, $d_{31}$, $d_{33}$ and $d_{51}$ are known as the piezoelectric coefficients found in the wurtzite crystal structure.[7] In any 3D coordinate system, each mutually independent element in the piezoelectric third rank tensor depends only on the crystallographic structure and orientations of the material. In the converse piezoelectric effect, an electric field is applied across the crystal generating strain or deformation in the material.[20] The piezoelectric tensor in AlGaN with space group of *P6₃mc*, has three independent constants, $d_{13}$, $d_{15}$, and $d_{33}$. It is reported that, the relation between strain-electric field for such kind of piezoelectric material with *6mm* point group symmetry ($\varepsilon_{ij} = d_{ijk}E_k$) can be expressed as follows. [15,20]

$$\begin{pmatrix} \varepsilon_{11} \\ \varepsilon_{22} \\ \varepsilon_{33} \\ 2\varepsilon_{23} \\ 2\varepsilon_{13} \\ 2\varepsilon_{12} \end{pmatrix} = \begin{pmatrix} 0 & 0 & d_{31} \\ 0 & 0 & d_{31} \\ 0 & 0 & d_{33} \\ 0 & d_{15} & 0 \\ d_{15} & 0 & 0 \\ 0 & 0 & 0 \end{pmatrix} \begin{pmatrix} E_1 \\ E_2 \\ E_3 \end{pmatrix} \quad \text{-------------------------------------------------- (5)}$$

From the above equation, the induced strain in presence of the external electric field along different directions is therefore expressed as $\varepsilon_{11} = \varepsilon_{22} = d_{31}E_3$, $\varepsilon_{33} = d_{33}E_3$, $2\varepsilon_{23} = d_{15}E_2$, $2\varepsilon_{13} = d_{15}E_1$, and $\varepsilon_{12} = 0$. Using the PFM set up, the displacement of cantilever due to the total polarization of the crystal can be measured as in terms of piezoelectric coefficient,

$$\Delta Z = d_{33}V_{dc} + d_{33}V_{ac}\cos(\omega t + \varphi), \quad \text{------------------------------------------ (6)}$$

The magnitude of the oscillating response is a measure of the magnitude of $d_{33}$ and the phase is sensitive to the polarization direction of the sample.[20] The vibrational amplitude is related to the vertical projection of the polarization. The vibration phase lag varies in between -180° to +180°, depending on whether the polarization is pointing up or down or the angular inclination between the external field and the polarization.

Figure 9 shows topography as well as PFM magnitude images of AlGaN microrods on AlN base layer over the intrinsic Si(100) substrate. The possible reason for contrast variation at the interface in the PFM images [Figs. 9(b) and 9(c)] may be due to the variation of surface charge density ($\sigma$) at the bottom surface of micro-crystals originating from the variation of crystalline orientation with respect to the AlN base layer. The surface charge density and immobile piezoelectric polarization charges may play an important role in the interaction of the wurtzite-structured AlGaN with the external electric field as well as the mechanical strain.[16] At the $Al_xGa_{1-x}N/AlN$ interface the surface charge density follows the relation,

$$\sigma = \nabla.P = x(P^{SP}_{AlN} - P^{SP}_{GaN})\cos\varphi \quad \text{------------------------------------------------------ (7)}$$

where, $\varphi$ is the angle of the interface between AlN base layer and AlGaN microrods with respect to the [001] direction. Therefore, angular orientation of the AlGaN micro-crystals with respect to the AlN base layer will also influence the piezoelectric polarization. It may lead to the variation of induced surface charge density over the micro-crystals and hence the different contrast in the PFM images.

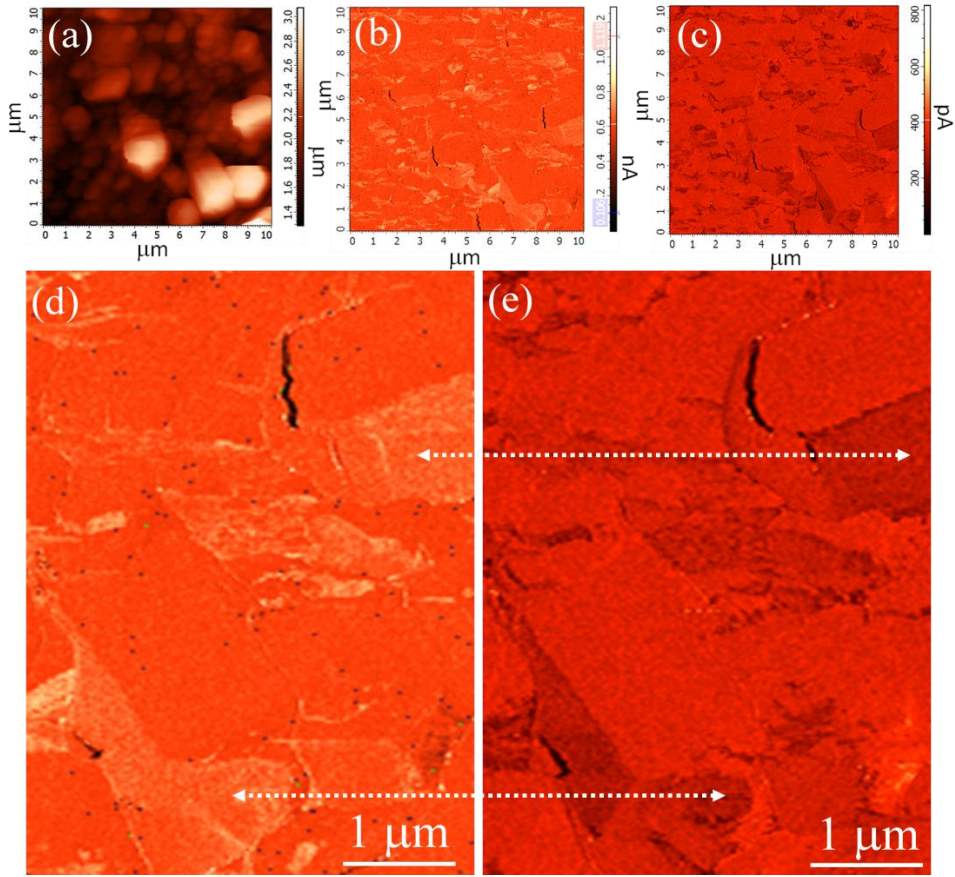

FIG. 9. (a) AFM topography of AlGaN hexagonal microrods and corresponding PFM images with an external voltage (b) +5V and (c) -5V. The domain switching in the piezoelectric zones indicated by arrows in the zoomed portion of PFM images of +5V and -5V in the (d) and (e), respectively.

The bright and dark contrast, corresponding domain expansion or contraction, is interchanging while switching the polarity of the external bias voltage from +5V [Fig. 9(d)] to -5V [Fig. 9(e)]. The dotted arrows in the figure indicate some specific area showing switching (flipping in) of the piezo-polarization with respect to the external bias voltage for different planes of the crystals. We also carried out the similar PFM experiment with a reduced external bias voltage of +2.5 and -2.5; the results are provided in the supplementary figure S7.[24] In piezo-spectroscopy mode, probing of electromechanical response can be carried out as a function of DC bias of the tip from a single point of the crystal (Fig. 10). The graph plotted between external bias voltages against phase change, represents how the polarity and magnitude of the external bias influence the phase change of the spontaneous electric polarization, $P^{SP}$. After application of one complete cycle of +ve to –ve external bias voltage, it is clearly visible that, the phase angle switching between the -20° to +20° even for a single spot [marked in topographs in Figs. 10 (a) and 10 (b)] on the AlGaN crystal itself. This explains and ensures a supplementary evidence for the switching behavior of polarization in single crystal AlGaN microrod.

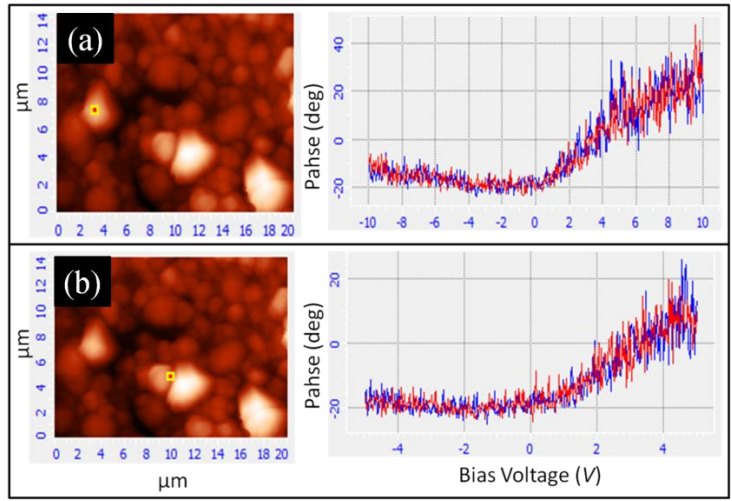

FIG. 10. The piezo-spectroscopy to shows the topographs and phase profiles from a single point on the AlGaN microrod with respect to a complete cycle of +ve to –ve external bias voltage for two different places of (a) and (b) of the film.

### *2. The piezoelectric coefficient ($d_{33}$) of AlGaN microrod*

The electric field, created along the microrod (*z*-direction) through the piezoelectric effect, $E_z = \varepsilon_{33}/d_{33}$. Therefore, the effective piezoelectric coefficient along the *c*-axis can be calculated as the ratio between deformation of the crystal (ΔZ) along *c*-axis and applied voltage *V*:

$$d_{33} = \varepsilon_{33}/E_z = \Delta Z/V \text{ -------------------------------------------- (8)}$$

where $\varepsilon_{33} = \Delta Z/Z$ is the change of strain along the *c*-axis as shown in the figures 1(a) and 1(b), and $E_z = V/Z$ is the electric field along the *c*-axis.[14,39-41] The strain coefficient, $\varepsilon_{33}$ is calculated by analyzing the deformation of crystal from line scan data of the PFM image. By using the above expression, we calculated the approximate value of $d_{33}$ along the *c*-axis as 4.14 pm/V. The calculated value of $d_{33}$ is slightly different from the reported values 5.1 and 3.1 pm/V for unclamped AlN and GaN, respectively.[39,42] The small deviations in the value of $d_{33}$ may be due to the following reasons. In one of the possibilities, the $d_{33}$ value for Al doped GaN may be slightly higher than that of the undoped GaN synthesized by CVD. At the same time, the AlGaN crystals grown on AlN base layer over Si(100) substrate are clamped by the substrate. The substrate generates some constrains to the expansion and contraction of the synthesized materials and thus affects the measured strains. As a result, the measured value of $d_{33}$ is less than the true value of an unclamped AlN and more than that of the GaN samples.[39,42]

## IV. CONCLUSIONS

In the present study we synthesized hexagonal microrods of AlGaN via self catalytic vapor-solid mechanism in the atmospheric pressure chemical vapor deposition technique. The proposed growth mechanism is substantiated with the help of X-ray diffraction and X-ray photoelectron spectroscopic analyses. The Raman and photoluminescence studies indicate the formation of high optical quality single crystalline AlGaN microrods in the wurtzite phase. AlGaN random alloy formation is confirmed from the presence of two mode behavior of $E_2^H$ phonons in the Raman spectra. The significant blue shift in the band edge emission compared to pure GaN is also a strong confirmation for the presence of Al in as-prepared sample. The approximate Al percentage is estimated ~ 2-3 at% from the blue shift of $A_1(LO)$ mode as well as the band edge emission. In this report, we investigated the piezoresponse force microscopic (PFM) imaging of AlGaN hexagonal microrods. The reasons for the contrast variation in the PFM image can be because of the variation of surface charge density over the surface of AlGaN microrods originating due to the crystalline orientation with respect to its AlN base layer. The effective piezoelectric coefficient for AlGaN microrod along the *c*-axis ($d_{33}$) is found to be 4.14 pm/V.

**ACKNOWLEDGMENTS**

One of us (A.K.S.) acknowledges the Department of Atomic Energy for allowing him to continue research work. We thank S. Ilango and Sunitha Rajkumari of SND, IGCAR for their help in the XRD study. We are extending our acknowledgment to S. R. Polaki for recording FESEM images. We also thank Avinash Patsha, Kishore K. Madapu and Bonu Venkataramana of SND, IGCAR for their valuable suggestions and useful discussions.

# Supplementary information:

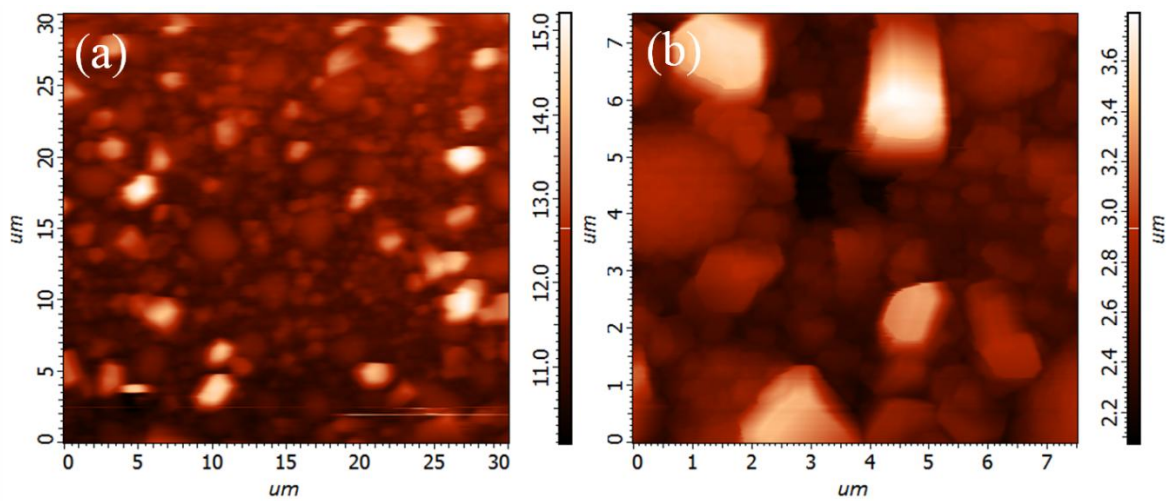

FIG. S1. The topographic image of AlGaN micro-rods with low (a) and high (b) magnifications.

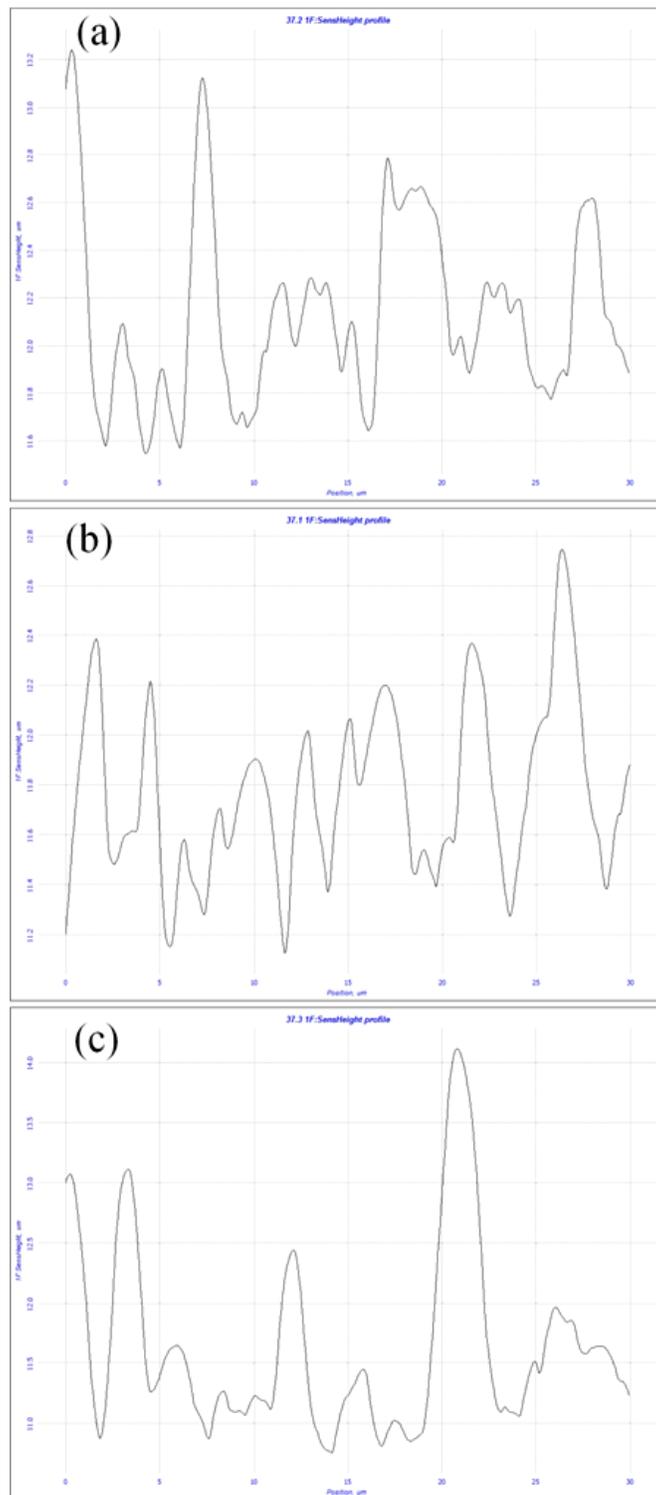

FIG. S2. (a), (b) and (c) are the line profiles of AlGaN mocro-rods along the X axis for different Y positions 25, 15 and 5 μm of the AFM topography shown in the figure S1(a). The RMS and average size distribution of AlGaN microrods found to be 0.634 and 0.440 μm, respectively.

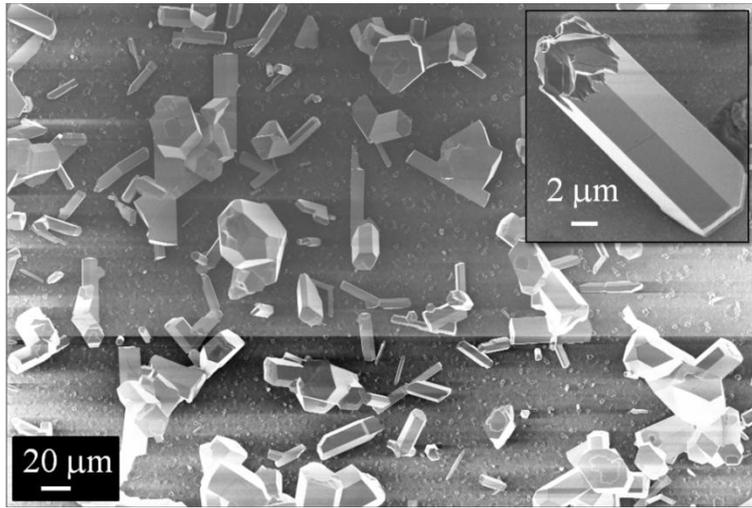

FIG. S3. The FESEM images of AlGaN hexagonal microrods grown on the sapphire substrate. Inset shows the magnified image of a single microrod.

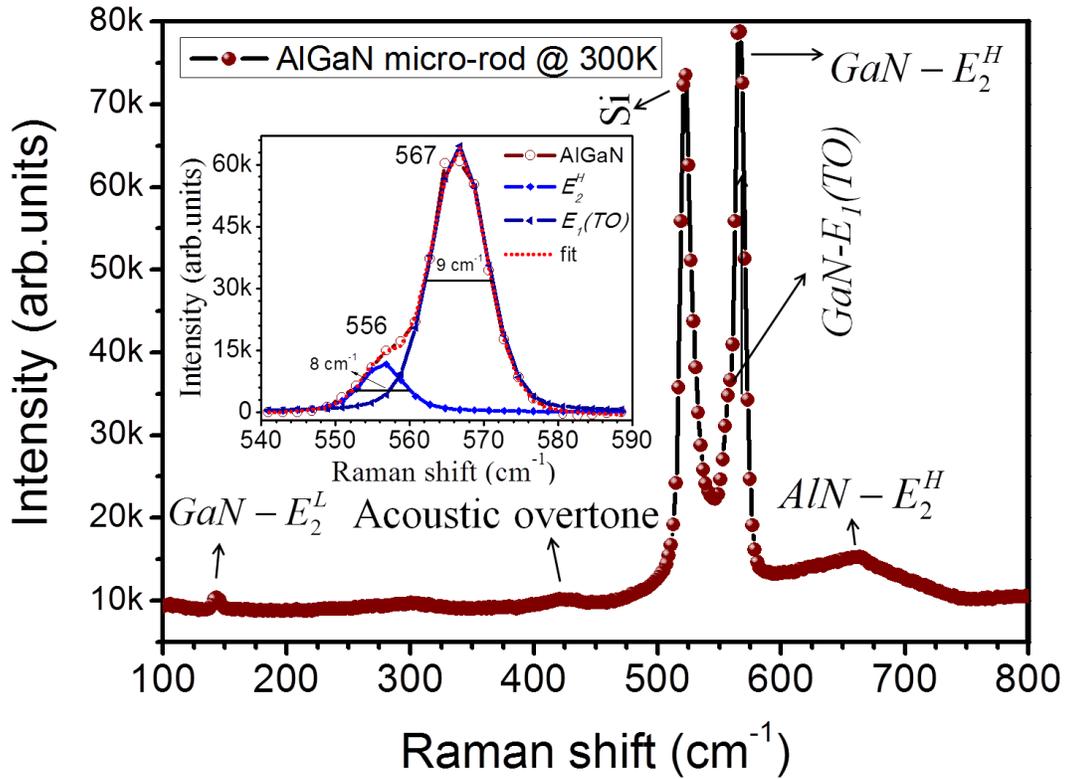

FIG. S4. Typical Raman spectrum (300K) of an AlGaN single crystal hexagonal microrod grown on sapphire substrate. Inset shows the Lorentzian fitted curve of $E_1(TO)$ and $E_2^H$ modes with FWHM of 8 and 9 cm$^{-1}$, respectively.

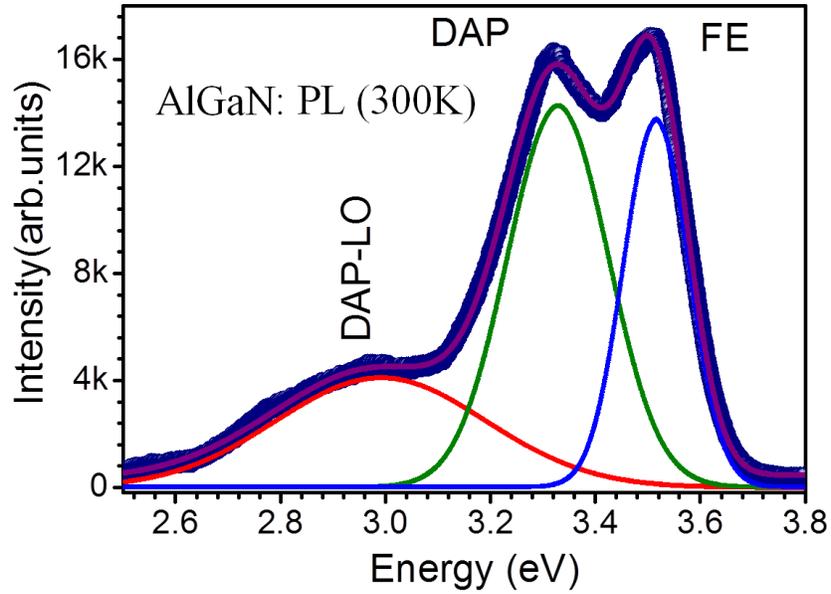

FIG. S5. Typical photoluminescence spectrum (300K) of an AlGaN single crystal hexagonal microrod grown on sapphire substrate.

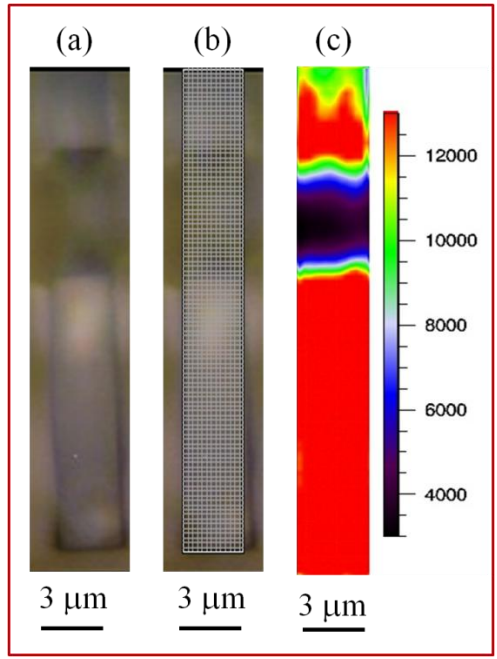

FIG. S6. (a) The optical image of an isolated AlGaN micro-rod (b) The rectangular area selected for Raman imaging of the crystal and (c) The integrated intensity distribution of $E_2^H$ mode along the micro-rod showing compositional homogeneity as well as crystalline nature.

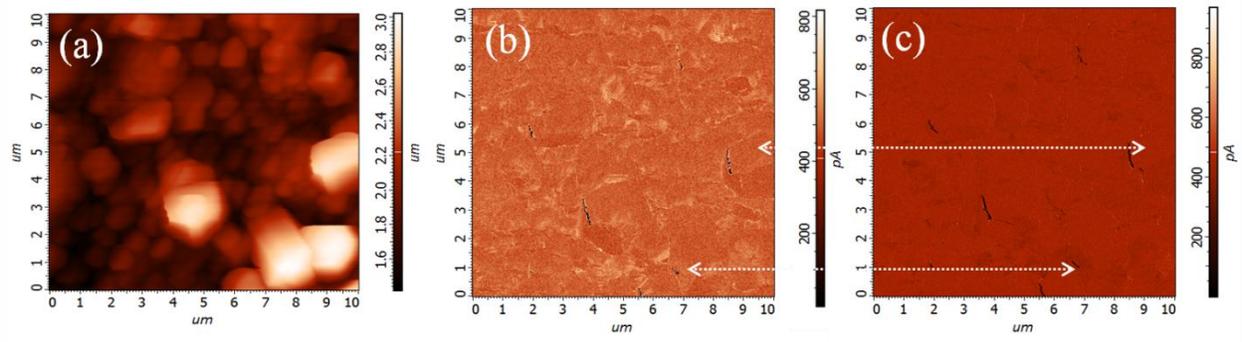

FIG. S7. (a) AFM topography of AlGaN hexagonal microrods and corresponding PFM images with an external voltage (b) +2.5V and (c) -2.5V. The domain switching in the piezoelectric zones indicated by arrows in the images.